\documentclass[10pt]{article}
\usepackage[abs]{overpic}
\usepackage{amsmath,amssymb,amsfonts}
\usepackage{graphicx,color}
\usepackage{setspace} 
\usepackage{hhline}

\makeatletter

\@addtoreset{equation}{section}
\makeatother

\newcommand{\be}{\begin{equation}}

\newcommand{\ee}{\end{equation}}
\newcommand{\bea}{\begin{eqnarray}}
\newcommand{\eea}{\end{eqnarray}}
\newcommand{\beann}{\begin{eqnarray*}}
\newcommand{\eeann}{\end{eqnarray*}}

\newcommand{\ba}{\begin{array}}
\newcommand{\ea}{\end{array}}
\newcommand{\tr}{\mathop{\rm tr}}

\newcommand{\N}{{\cal N}}
 
\def\XXint#1#2#3{{\setbox0=\hbox{$#1{#2#3}{\int}$} 
\vcenter{\hbox{$#2#3$}}\kern-.5\wd0}} 
 
\renewcommand{\thefootnote}{\fnsymbol{footnote}}

\begin{document}

\setlength{\oddsidemargin}{0cm}
\setlength{\baselineskip}{7mm}
\vfil\eject
\setcounter{footnote}{0}
\begin{flushright}
\normalsize
OCU-PHYS 354\\
~\\
~\\
~\\
~\\
\end{flushright}
    \begin{Large}  
       \begin{center}
         { \bf Baxter's T-Q equation, $SU(N)/SU(2)^{N-3}$ correspondence and\\ $\Omega$-deformed Seiberg-Witten prepotential
}
       \end{center}
        \end{Large}
        
\begin{center}
Kenji Muneyuki,$ ^a$\footnote[1]{
        e-mail address: 1033310118n@kindai.ac.jp
    }
Ta-Sheng Tai,$ ^a$\footnote[2]{
        e-mail address: tasheng@alice.math.kindai.ac.jp
    }
Nobuhiro Yonezawa$ ^b$\footnote[3]{
        e-mail address: yonezawa@sci.osaka-cu.ac.jp
    }
and
Reiji Yoshioka$ ^b$\footnote[4]{
        e-mail address: yoshioka@sci.osaka-cu.ac.jp
    }
\end{center}
\renewcommand{\thefootnote}{\arabic{footnote}}
\begin{small}
\begin{center}
{\it $\ ^a$Interdisciplinary Graduate School of Science and Engineering,\\
Kinki University, 3-4-1 Kowakae, Higashi-Osaka, Osaka 577-8502, Japan}\\
\end{center}
\begin{center}
{\it $\ ^b$Osaka City University Advanced Mathematical Institute,\\
3-3-138 Sugimoto, Sumiyoshi-ku, Osaka 558-8585, Japan} \\
\end{center}
 \end{small}
\begin{abstract}
\noindent
{\normalsize
We study Baxter's T-Q equation of XXX spin-chain models under the semiclassical limit where an 
intriguing $SU(N)/SU(2)^{N-3}$ correspondence emerges. That is, 
two kinds of 4D $\N=2$ superconformal field theories having the above different gauge groups are encoded simultaneously 
in one Baxter's T-Q equation which captures their spectral curves. 
For example, 
while one is 
$SU(N_c)$ with $N_f=2N_c$ flavors the other turns out to be 
$SU(2)^{N_c-3}$ with $N_c$ 
hyper-multiplets ($N_c > 3$). 
It is seen that the corresponding Seiberg-Witten differential supports our proposal. \normalsize
}
\end{abstract}
\setstretch{1.1}
\section{Introduction and summary}
Recently, there have been new insights into the 
duality between integrable systems and 4D $\N=2$ gauge theories. 
In \cite{NS:0901,NS:0908,Nekrasov:2011bc} Nekrasov and Shatashvili (NS) have found that Yang-Yang functions 
as well as Bethe Ansatz equations of a family of integrable models are indeed encoded in a variety of Nekrasov's partition functions \cite{N1,N2} restricted to the 
two-dimensional $\Omega$-background%
\footnote{See also recent \cite{Orlando:2010aj, Zenkevich:2011zx, Dorey:2011pa, Chen:2011sj, Gaiotto_Witten_2011} which investigated XXX spin-chain models along this line.}. As a matter of fact, this mysterious correspondence 
can further be extended to the full $\Omega$-deformation in view of the birth of AGT conjecture \cite{Alday:2009aq}. 
Let us briefly refine the latter point. 
\begin{figure}[h]
  \begin{center}
    \includegraphics{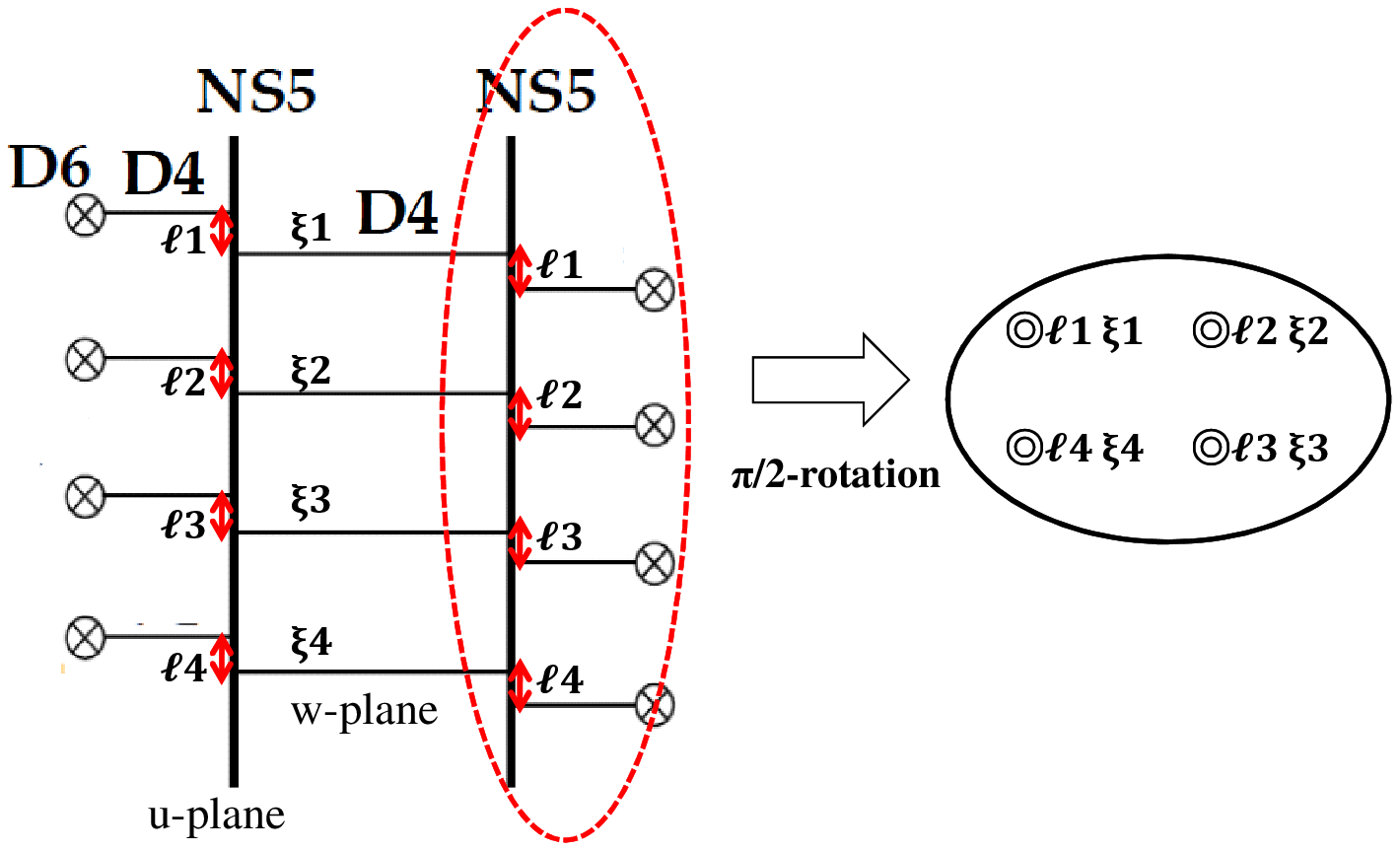}
  \end{center}
  \caption{Main idea: $SU(N)/SU(2)^{N-3}$ correspondence \protect\\
LHS: M-theory curve of $SU(4)$ $N_f=8$ Yang-Mills theory embedded in 
$\bf{C} \times \bf{C}^\ast$ parameterized by 
$(u,w)$ $(w=\exp(-s/R),~R=\ell_s g_s:~\text{M-circle~radius})$; RHS: spin-chain variables $(\xi_n, \ell_n)$ labeling 
(coordinate, weight) of each puncture on 
${\bf{CP}}^1$ (but indicating each flavor D6-brane location along $u$-plane of LHS) }
\label{2}
\end{figure}

AGT claimed that 
correlators of 
primary states in 
Liouville 
field theory (LFT) can 
get re-expressed in terms of Nekrasov's partition function 
$Z_{\text{Nek}}$ 
of 4D ${\cal N}=2$ quiver-type $SU(2)$ superconformal field theories (SCFTs). 
In particular, every 
Riemann surface $C_{g,n}$ (whose doubly-sheeted cover is 
called Gaiotto curve \cite{D_Gaiotto_0904}) on which LFT dwells 
is responsible for one specific SCFT called 
${\cal T}_{g,n}(A_{1})$ such that 
the following equality 
\begin{eqnarray*}
\text{conformal~block~w.r.t.~}C_{g,n} = 
 \text {instanton~part~of}~Z_{\text{Nek}}\Big({\cal T}_{g,n}(A_{1})\Big)
\end{eqnarray*}
holds. Because of $\epsilon_{1}:\epsilon_{2}=b^{-1} : b$ 
\cite{Alday:2009aq} the one-parameter version ($\epsilon_{2}=0$) of AGT conjecture directly leads to the semiclassical 
LFT as $b\to 0$. Quote further the geometric Langlands 
correspondence \cite{FFR} which associates Gaudin integrable 
models on the projective line with LFT at $b\to 0$. It is then plausible to put both proposals of NS and 
AGT into one unified scheme.

In this letter, we add a new element into the above 2D/4D correspondence. 
Starting from 
Baxter's T-Q equation of XXX spin-chain models we found a novel interpretation of it. That is, under the semiclassical limit it 
possesses two aspects simultaneously. It describes \\
$\bullet$ 4D $\N=2$ 
$SU(N_c)$ Yang-Mills with $N_f=2N_c$ flavors, ${\cal T}_{0,4}(A_{N_c -1})$, on the one hand and \\
$\bullet$ $SU(2)^{N_c-3}$ ($N_c > 3$) quiver-type Yang-Mills with $N_c$ 
(four fundamental and $N_c-4$ bi-fundamental) hyper-multiplets, 
${\cal T}_{0,N_c}(A_{1})$, on the other hand. \\It is helpful to have a rough 
idea through Fig. \ref{2}. Pictorially, $C_{0,4}$ for 
${\cal T}_{0,4}(A_{1})$ in RHS results from the encircled part 
in LHS after a 
$\pi/2$-rotation. 
\begin{table}[h]
\begin{center}
\begin{tabular}{c|c|c|c|c}
\hline
&\raisebox{-1pt}{0, 1, 2, 3}&\ \raisebox{-1pt}{$u=x^4 + i x^5$}\ \ &\ \raisebox{-1pt}{6}\ \ &\raisebox{-1pt}{7, 8, 9} \\ \hhline{=|=|=|=|=}
\raisebox{-1pt}{D6}&$\circ$& - &-& $\circ$     \\ \hline
\raisebox{-1pt}{NS5}& $\circ$ & $\circ$ &-&-     \\ \hline
\raisebox{-1pt}{D4} & $\circ$ & -&$\circ$&- \\ \hline
\end{tabular}
\caption{Type IIA D6-NS5-D4 brane configuration}
\label{tab1}
\end{center}
\end{table}
In other words, the conventional Type IIA Seiberg-Witten (SW) curve (see Table \ref{tab1}) in fact contains another important 
piece of information while seen from $(u,v)$-space ($u=x^4 + ix^5, v=x^7 + i x^8$)%
\footnote{This aspect of $\N=2$ curves is also stressed in \cite{Ohta:2008md}. }. 
Here, ``$\pi/2$-rotation" just means that SW differentials of two theories thus yielded are connected by 
exchanging $(u, s=x^6 + ix^{10})$.

This quite unexpected phenomenon 
will be explained later by combining a couple of topics, say, Bethe Ansatz, Gaudin model and Liouville theory. 
Roughly speaking, the spin-chain variable $\ell$ $(\xi)$, 
highest weight (shifting parameter), is responsible for $m$ $(q)$ of RHS in Fig. \ref{2}. 
As summarized in Table \ref{tab2}, $N_c$ Coulomb moduli $\xi\in \bf{C}$ 
(one overall $U(1)$ factor) are mapped to $N_c-3$ 
gauge coupling constants $q=\exp(2\pi i\tau)\in \bf{C}^\ast$ where three of them are fixed to $(0, 1, \infty)$ on $\bf{C}^\ast$. 
Those entries marked by $\odot $ do not have direct comparable counterparts. 
\begin{table}[ht]
\begin{center}
\begin{tabular}{ccc}
\hline
$\#$ of UV parameter& LHS&RHS\\ \hline
Coulomb moduli&$N_c-1~({\xi})$& $\odot  N_c -3$~($a$) \\ \hline
bare flavor mass &$N_c~({\xi \pm \ell })$& $N_c~({m})$ \\ \hline
gauge coupling & $\odot$1~$\exp(\dfrac{\Delta x^6 + i\Delta x^{10}}{R})$ &  $N_c-3~(q)$ \\
\hline
\end{tabular}
\caption{UV parameters of two $\N=2$ theories ($N_c > 3$) in Fig. \ref{2} }
\label{tab2}
\end{center}
\end{table}

We organize this letter as follows. Sec. 2 is 
devoted to a further study of Fig. \ref{11} on which our main idea Fig. \ref{2} is based. 
Then Sec. 3 unifies three elements: Gaudin model, LFT and matrix model as shown in Fig. \ref{3}. 
Finally, in Sec. 4 we complete our proposal by examining $\lambda_{SW}$ (SW differential) and shortly discuss XYZ Gaudin models.


\section{XXX spin chain}
 
Baxter's T-Q equation \cite{B1:1972,B2:1973} plays 
an underlying role in various spin-chain models. On the other hand, it has long been known that 
the low-energy Coulomb sectors of $\N=2$ gauge theories are intimately related to a variety of integrable systems \cite{1,2,3,Itoyama:1995nv,Itoyama:1995uj,4,5}. Here, by integrable model (or solvable model) we mean that there exists some spectral curve which gives enough integrals of motion (or conserved charges). 
In the case of $\N=2$ $SU(N_{c})$ Yang-Mills 
theory with $N_{f}$ fundamental hyper-multiplets, its SW curve \cite{01,02} 
is identified with the spectral curve of an inhomogeneous periodic Heisenberg XXX spin chain 
on $N_c$ sites: 
\begin{eqnarray}
w+ \frac{1}{w}=\frac{P_{N_c}(u)}{\sqrt{Q_{N_f}(u)}}.
\label{1}
\end{eqnarray}
Here, two 
polynomials $P_{N_{c}}$ and $Q_{N_{f}}$ encode respectively parameters of 
$\N=2$ vector- and hyper-multiplets. Meanwhile, 
the meromorphic SW differential $\lambda_{SW}=ud\log w$ provides a 
set of ``special coordinates" through its period integrals (see Table \ref{tab2}): 
\begin{eqnarray}
\xi_n =\oint_{\alpha_n} \lambda_{SW}, ~~~~~~
\frac{\partial {\cal F}_{SW}}{\partial \xi_n}=\xi^D_n =\oint_{\beta_n} \lambda_{SW}, ~~~~~~
\xi_n \pm \ell_n=\oint_{\gamma_n^{\pm}} \lambda_{SW}
\label{phypre}
\end{eqnarray}
where ${\cal F}_{SW}$ is the physical prepotential.

\subsection{Baxter's T-Q equation}
Indeed, \eqref{1} arises from (up to $w\to \sqrt{Q_{N_f}}w$)
\begin{eqnarray*}
&&\det\big(w-{T}(u)\big)=0~~\to~~w^{2}-\tr T(u)w+\det T(u)=0, ~~~~~~T(u):~2\times 2\text{~monodromy~matrix},\\
&&\det T(u)=Q_{N_{f}}(u)=\prod_{n=1}^{N_c} (u-m^-_n)(u-m^+_n), ~~~~~~m^{\pm}_n=\xi_n\pm\ell_n. 
\end{eqnarray*}
$\tr T(u)=t(u)=P_{N_{c}}(u)=\langle \det (u-\Phi)\rangle$, transfer matrix, encodes the quantum vev of the adjoint scalar field $\Phi$. In fact, \eqref{1} belongs to the $conformal$ case where 
$N_f=2N_c$ bare flavor masses are indicated by $m_n^{\pm}$. It is time to quote Baxter's T-Q equation:
\begin{eqnarray}
t(u)Q(u)=\triangle_{+}(u)Q(u-2\eta)+\triangle_{-}(u)Q(u+2\eta).
\label{pretq}
\end{eqnarray}
Some comments follow: \\
$\bullet$ $\eta$ is Planck-like and ultimately gets identified with $\epsilon_{1}$ 
(one of two $\Omega$-background parameters)  in Sec. 4. \\
$\bullet $ As a matter of fact, \eqref{pretq} boils down to \eqref{1} (up to $w\to \sqrt{Q_{N_f}}w$) as $\eta\to 0$. 
Curiously, then its $\lambda_{SW}$ signals the existence of another 
advertised ${\cal N}=2$ theory. The situation is 
pictorially shown in Fig. \ref{2}. \\
$\bullet $ Remark again that SW differentials of two theories are 
connected by exchanging two holomorphic coordinates 
$(u, s)$ but their M-lifted \cite{Witten:1997sc} 
Type IIA brane configurations%
\footnote{In \cite{Gorsky:1997jq} 
this symmetry has been notified in the context of Toda-chain models because two kinds of Lax matrices exist there.} 
are not. Instead, the $\pi/2$-rotated part is closely related to 
${\cal N}=2$ Gaiotto's curve. A family of quiver-type $SU(2)$ SCFTs ${\cal T}_{0,n}(A_1)$ 
discovered by 
Gaiotto \cite{D_Gaiotto_0904} is hence made contact with. 

\begin{figure}[ht]
  \begin{center}
    \includegraphics{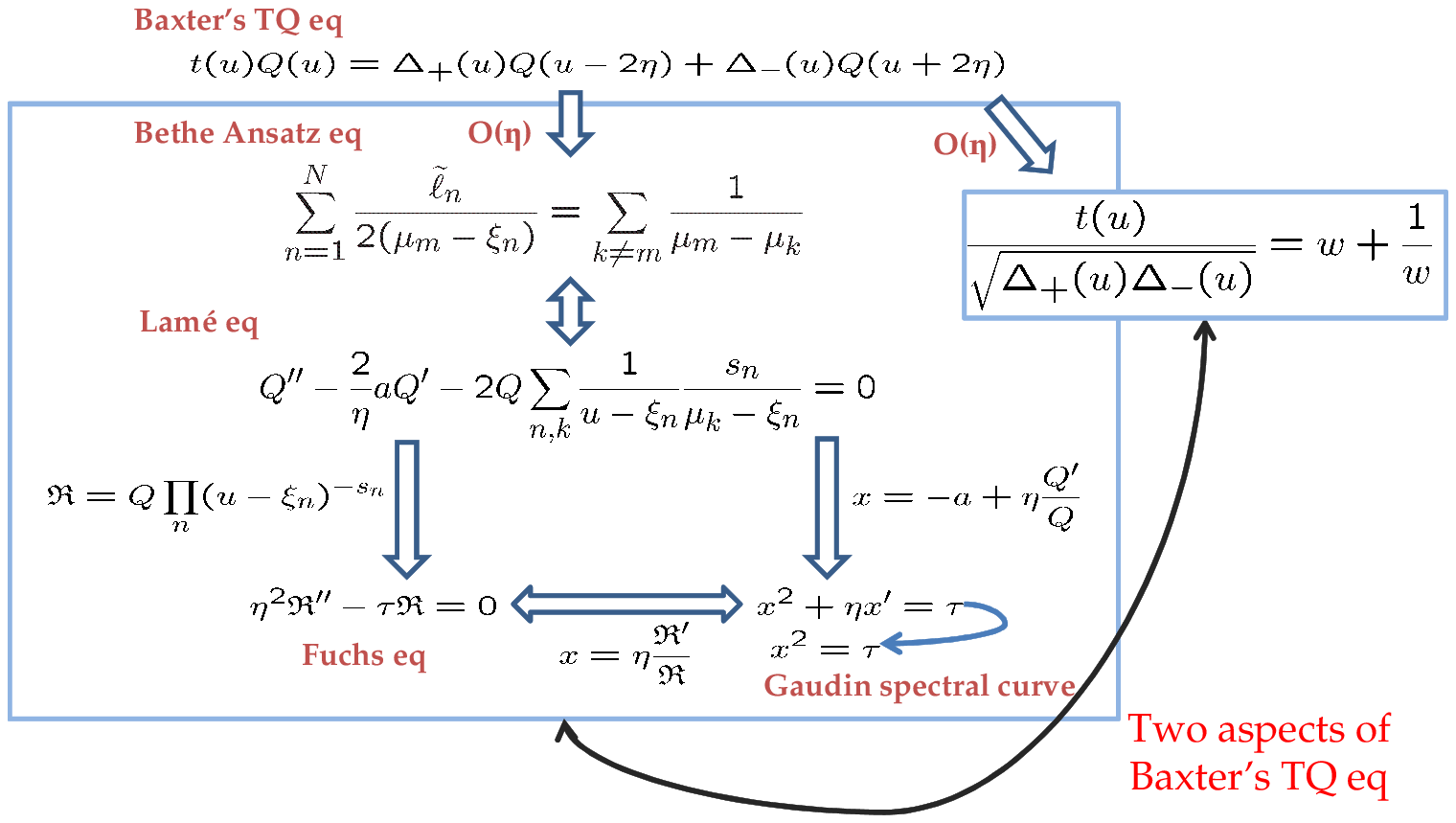}
  \end{center}
  \caption{Mathematical description of Fig. \ref{2} \protect\\
Up to ${\cal O}(\eta)$, Baxter's T-Q equation and Bethe Ansatz 
equations of it describe two kinds of ${\cal N}=2$ gauge theories which however are related by one $\lambda_{SW}$ }
    \label{11}
\end{figure}

\subsection{More detail}
Let us refine the above argument. 
Consider a quantum spin-chain built over an $N$-fold tensor product ${\cal H}=\otimes_{n=1}^{N}V_{n}$. 
In other words, at each site labeled by $n$ we assign an irreducible 
representation $V_{n}$ of ${\mathfrak{sl}}_2$ which 
is $(\ell_{n}+1)$-dimensional where $\ell_{n}=0,1,2,\cdots$. 
Therefore, $\ell_{n}$ denotes the highest weight. Within 
the context of QISM%
\footnote{Quantum Inverse Scattering Method (QISM) was formulated in 1979-1982 
in St. Petersburg Steklov Mathematical Institute by Faddeev and many of 
his students. 
We thank Petr Kulish for informing us of this fact. }, monodromy and transfer 
matrices are defined respectively by 
\begin{eqnarray}
&&T(u)=\left( 
\begin{array}{cc}
A_{N}(u)&B_{N}(u)\\C_{N}(u)&D_{N}(u)
\end{array}\right)=L_{N}(u-\xi_{N})\cdots L_{1}(u-\xi_{1}),\\
&&\widehat{t}(u)=A_{N}(u)+D_{N}(u).
\nonumber\label{first}
\end{eqnarray}
$V_{n}$ is acted on by the $n$-th Lax operator $L_{n}$. 
By $inhomogeneous$ one means that the spectral parameter $u$ has been shifted by $\xi$. 
Conventionally, 
$\widehat{t}(u)$ or its eigenvalue $t(u)$ is called generating function because 
a series of conserved charges can be extracted from its coefficients owing to $[\widehat{t}(u),\widehat{t}(v)]=0$. 
The commutativity arises just from the celebrated Yang-Baxter equation.

As far as 
the inhomogeneous periodic XXX spin chain is concerned, its T-Q equation reads ($\ell=\eta\widetilde{\ell}$)
\begin{eqnarray}
&&t(u)Q(u)= \triangle_{+}(u)Q(u-2\eta)+\triangle_{-}(u)Q(u+2\eta),\label{eq_T-Q_eq}\nonumber\\
&&Q(u)=\prod_{k=1}^{K}(u-\mu_{k}),~~~~~~
\triangle_{\pm}=\prod_{n=1}^{N}{(u-\xi_{n}\pm \eta \widetilde{\ell}_n)}
\label{tq}
\end{eqnarray}
where each Bethe root $\mu_{k}$ satisfies a set of Bethe Ansatz equations: 
\begin{eqnarray}
\frac{\triangle_{+}(\mu_{k})}{\triangle_{-}(\mu_{k})}=
\prod_{n=1}^{N}\frac{{(\mu_{k}-\xi_{n}+\eta \widetilde{\ell}_n)}}{{(\mu_{k}-\xi_{n}-\eta \widetilde{\ell}_n)}}=
\prod_{l(\ne k)}^{K}\frac{\mu_{k}-\mu_{l}+2\eta}{\mu_{k}-\mu_{l}-2\eta}.\label{b}
\end{eqnarray}
A semiclassical 
limit is facilitated by the $\eta$ dependence. Through 
\begin{eqnarray}
\frac{t(u)}{\sqrt{\triangle_{+} \triangle_{-}}}=
\frac{Q(u-2\eta)}{Q(u)}
\sqrt{\frac{\triangle_{+}}{\triangle_{-}}}+
\frac{Q(u+2\eta)}{Q(u)}
\sqrt{\frac{\triangle_{-}}{\triangle_{+}}}
\end{eqnarray}
and omitting ${\cal O}(\eta^2)$, we have 
\begin{eqnarray}
\frac{t(u)}{\sqrt{\triangle_{+}(u) \triangle_{-}(u)}}=
w+ \frac{1}{w},~~~~~~~~~
w\equiv \sqrt{\frac{\triangle_{+}}{\triangle_{-}}}(1-2\eta \frac{Q'}{Q})
\label{new}
\end{eqnarray}
which exactly reduce to \eqref{1}. Throughout this letter, 
($\eta, \widetilde{\ell}$) while kept finite are, respectively, small and large.

From now on, we call $\lambda_{SW}= ud\log w\equiv \lambda_{SW}^{\eta}$ 
``$\eta$-deformed" SW differential as in 
\cite{Poghossian:2010pn, Fucito:2011pn}: 
\begin{eqnarray}
\lambda_{SW}^{\eta}= 2\eta u 
d\left(\frac{\Psi^{\prime}}{\Psi}\right)+{\cal O}(\eta^{2})
,~~~~~~~~~\Psi=\frac{1}{Q(u)}\prod_{n}
(u-\xi_{n})^{\widetilde{\ell}_{n}/2}.
\label{swd}
\end{eqnarray}
Also, up to ${\cal O}(\eta)$ 
\eqref{b} reads
\begin{eqnarray}
\sum_{n=1}^{N}\frac{\widetilde{\ell}_{n}}{2(\mu_{k}-\xi_{n})}=
\sum^K_{l(\ne k)}\frac{1}{(\mu_{k}-\mu_{l})}.
\label{cb}
\end{eqnarray}
That $\lambda^\eta_{SW}$ looks strikingly similar to \eqref{cb} signals the existence of RHS in Fig. \ref{2}. 
Fig. \ref{11} outlines our logic. One will find that 
$\lambda_{SW}^{\eta}$ naturally emerges as the 
holomorphic one-form of Gaudin's spectral curve 
which captures 
Gaiotto's curve for ${\cal T}_{0,N}(A_{1})$. 
In what follows, our goal is to show that 
$\lambda_{SW}^{\eta}$ does reproduce the $\epsilon_1$-deformed SW 
prepotential w.r.t. 
${\cal T}_{0,N}(A_{1})$.

Several comments follow: \\
$\bullet $ In M-theory 
D6-branes correspond to singular loci of 
$XY ={{\triangle_{+}(u) \triangle_{-}(u)}}$. This simply 
means that one incorporates flavors via 
replacing a flat ${\mathbf{R}}^4$ over $(u,s)$ by a resolved ${\bf A}_{2N_c -1}$-type singularity. \\
$\bullet $ Without flavors (i.e. turning off $\ell$) 
$\int \lambda_{SW}^{\eta}$ looks like a logarithm of the usual Vandermonde. 
This happens in the familiar Dijkgraaf-Vafa story \cite{D1,D2,D3} without any 
tree-level potential which brings $\N=2$ pure Yang-Mills 
to $\N=1$ descendants. \\
$\bullet $ Surely, this intuition is noteworthy in view of 
\eqref{cb} which manifests itself as the saddle-point condition within the context of matrix models. 
To pursue this interpretation, one should regard 
$\mu$'s as diagonal elements of ${\cal M}$ (Hermitian matrix of size $K \times K$). 
Besides, the 
tree-level potential now obeys 
\begin{eqnarray*}
{\cal W}'(x)=-\sum_{n=1}^{N}\frac{ \ell_{n}}{(x-\xi_{n})}.\label{cc}
\end{eqnarray*}
In other words, we are equivalently dealing with ``$\N=2$" Penner-type matrix models 
which have been heavily investigated recently in connection with AGT conjecture 
due to \cite{Dijkgraaf:2009pc}. We will return to these points soon.

\section{XXX Gaudin model}
Momentarily, we focus on another well-studied integrable model: 
XXX Gaudin model. The essential difference between Heisenberg and Gaudin models amounts to the definition of their 
generating functions. Following Fig. \ref{3} 
we want to explain two 
important aspects of Gaudin's spectral curve. 
\begin{figure}[ht]
  \begin{center}
    \includegraphics{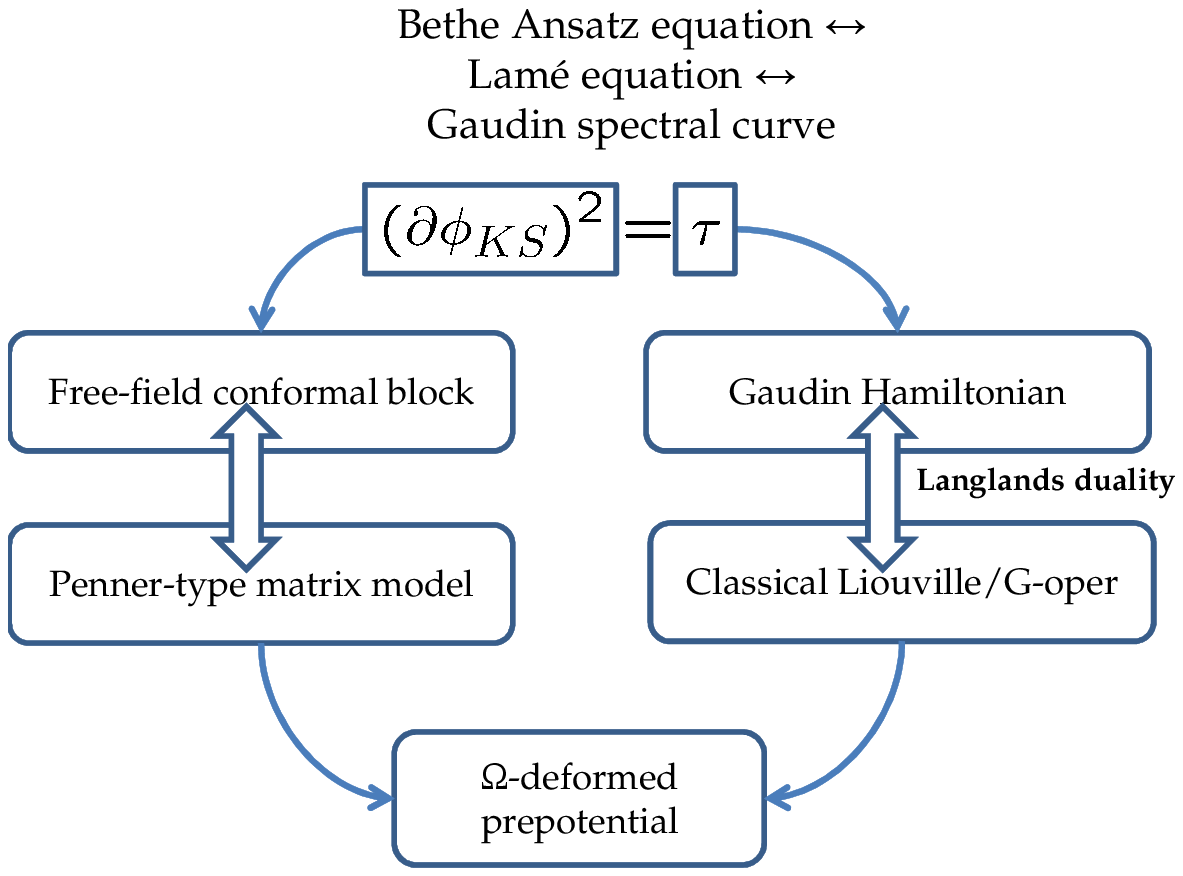}
  \end{center}
  \caption{Flow chart of Sec. 3}
    \label{3}
\end{figure}

\subsection{RHS of Fig. 3}

Expanding around small $\eta$, we yield 
\begin{eqnarray}
&&L_{n}(u)=1+2\eta{\cal L}_n+O(\eta^{2}),\\&& T(u)=1+2\eta{\cal T}+\eta^{2}{\cal T}^{(2)}+O(\eta^{3}),\\&& t(u)=1+\eta^{2}{\text{tr}}{\cal T}^{(2)}+O(\eta^{3}),\\
&&\tau(u)\equiv \frac{1}{2}\eta^2{\text tr}{\cal T}^2, 
~~~~~~~~~~{\cal T}=\sum_n {\cal L}_n=\left(\begin{array}{cc}A(u)&B(u)\displaystyle \\C(u)&-A(u)\displaystyle \end{array}\right)
\end{eqnarray}
where
\begin{eqnarray}
A(u)=\displaystyle \sum_{n=1}^{N}\frac{J_{n}^{z}}{u-\xi_{n}},~~~~~~~~~B(u)=\sum_{n=1}^{N}\frac{J_{n}^{-}}{u-\xi_{n}},~~~~~~~~~C(u)=\sum_{n=1}^{N}\frac{J_{n}^{+}}{u-\xi_{n}}
\end{eqnarray}
 while $\vec{J}=(J^{z},J^{\pm})$ represents generators of ${\mathfrak{sl}}_2$ Lie algebra. 
Instead of $\tr{\cal T}^{(2)}$ ($\tr{\cal T}=0$) the generating function 
adopted is 
($s=\widetilde{\ell}/2=\ell/2\eta$)
\begin{eqnarray}
\tau(u)=
\sum_{n=1}^{N}\Big\{ \frac{ \eta^2 s_n(s_n+1)}{(u-\xi_{n})^{2}}
+\frac{c_{n}}{u-\xi_{n}}\Big \},
~~~~~~~~
~c_{n}=\sum_{i\ne n}^{N}\frac{2 \eta^2 \vec{J}_{n}\cdot\vec{J}_{i}}{\xi_{n}-\xi_{i}},
~~~~~~~~~\vec{J}_{n}\cdot\vec{J}_{n}= s_n(s_n+1).
\end{eqnarray}
Conventionally, $c_{n}$'s are called Gaudin Hamiltonians which commute with one another as a result of the classical Yang-Baxter equation. 
\begin{eqnarray*}
\Sigma:~x^2=\tau(u) \subset {T^\ast C}
\end{eqnarray*}
is the $N$-site Gaudin spectral curve, a 
doubly-sheeted cover of $C\equiv {\bf{CP}}^1 \backslash\{\xi_{1},\cdots,\xi_{N}\}$.

According to the geometric Langlands correspondence%
\footnote{See also \cite{Teschner:2010je,Frenkel:2005pa, Tai:2010gd}.}, 
$c_{n}$'s give exactly $accessory$ parameters of a $G$-oper:  
\begin{eqnarray*}
{\cal D}=-\displaystyle \partial_{z}^{2}+
\sum_{n=1}^{N}
\frac{\delta_{n}}{(z-\xi_{n})^{2}}+\sum_{n=1}^{N}
\frac{\widetilde{c}_{n}}{z-\xi_{n}},
~~~~~~~~~
\delta=s(s+1),
~~~~~~~~~
c=\eta^2 \widetilde{c}
\label{pgl}
\end{eqnarray*}
defined over $C={\bf{CP}}^1 \backslash\{\xi_{1},\cdots,\xi_{N}\}$. 
The non-singular behavior of ${\cal D}$ is ensured by imposing 
\begin{eqnarray*}
\sum_{n=1}^N {\widetilde{c}}_{n}=0,~~~~~~~~~
\sum_{n=1}^{N}(\xi_{n}\widetilde{c}_{n}+\delta_{n})=0,~~~~~~~~~
\sum_{n=1}^{N}(\xi_{n}^{2}\widetilde{c}_{n}+2\xi_{n}\delta_{n})=0.
\end{eqnarray*}

Certainly, one soon realizes that $\tau(u)$ here is nothing but 
the holomorphic LFT (2,0) stress-tensor 
as the central charge $1+6Q^{2}$ goes to infinity (or $b\to 0$). 
Namely, 
\begin{eqnarray*}
\eta^{-2}\tau \equiv \frac{1}{2}\partial_{z}^{2}\varphi_{cl}-\frac{1}{4}(\partial_{z}\varphi_{cl})^{2}=\sum_{n=1}^{N}\frac{\delta_{n}}{(z-\xi_{n})^{2}}+\sum_{n=1}^{N}\frac{\widetilde{c}_{n}}{z-\xi_{n}}.
\end{eqnarray*}
In terms of LFT, the second equality comes from Ward identity of the stress-tensor $T_L= \frac{1}{2}Q 
\partial_{z}^{2}\phi
-\frac{1}{4}(\partial_{z}\phi)^{2}$ inserted in 
$\langle\prod_{n}V_{\alpha_{n}}\rangle$ subject to 
$b\to 0$. Here,  $V_{\alpha}=\exp({2}\alpha\phi)$ denotes the  primary field $(\Delta_{\alpha}=\alpha(Q-\alpha),~Q=b+{b}^{-1})$. As $b\to 0$, 
\begin{eqnarray*}
\big\langle  (-T_L) \prod_{n}V_{\alpha_{n}} \big\rangle =
\int {\cal D}\phi \exp(-{\bf S}_{\text{tot}}) (-T_L)~
\to ~\exp (-\dfrac{1}{b^2}{\cal S}_{\text{tot}}[\varphi_{cl}]) \dfrac{1}{\eta^2 b^2}\tau 
\end{eqnarray*}
such that for the unique saddle-point to ${\cal S}_{\text{tot}}[\varphi]$ 
one has (Polyakov conjecture)
\begin{eqnarray}
\widetilde{c}_n=  \frac{\partial{\cal S}_{\text{tot}}[\varphi_{cl}]}{\partial\xi_{n}},~~~~
~~~~~|\widetilde{\alpha}_n|=b|\alpha_n| =s_n  
\label{polyakov}
\end{eqnarray}
where on a large disk $\Gamma$ 
\begin{eqnarray*}
{\bf S}_{\text{tot}}=\int_{\Gamma} d^{2}z \Big(\frac{1}{4\pi} |\partial_{z}\phi|^{2}+ {\mu}e^{{2} b\phi} \Big)
+\text{boundary~terms},~~~~~~~~~{\bf S}_{\text{tot}}[\phi]=
\dfrac{1}{b^2}{\cal S}_{\text{tot}}[\varphi].
\end{eqnarray*}
Note that $\varphi_{cl}$ satisfies Liouville's equation and 
is important during uniformizing Riemann surfaces with constant 
negative curvature. 
Usually, $\widetilde{\alpha}=b\alpha$ is kept fixed 
during $b\to 0$. It is necessary that $\eta=\hbar /b$ due to 
$b|\alpha|= \ell/2\eta$. 
This confirms in advance $\eta =\epsilon_{1}$ due to AGT dictionary.

\subsection{LHS of Fig. 3}

As shown in \cite{Babelon_Talalaev_2007},  $\tau(u)$ has another form in terms of the eigenvalue $a(u)$ of $A(u)$%
\footnote{We hope that readers will not confuse $a(u)$ here with 
$a$ denoting Coulomb moduli.}:
\begin{eqnarray}
 \tau(u)=a^{2}-\eta a^{\prime}-2\eta\sum_{k}\frac{a(u)-a(\mu_{k})}{u-\mu_{k}},~~~~~~~~~a(u)\equiv \sum_{n=1}^{N}
\frac{\eta s_{n}}{(u-\xi_{n})}
\end{eqnarray}
with $ \mu_k$'s being Bethe roots. 
This expression is extremely illuminating in connection with Penner-type matrix models. 
Borrowing $Q(u)$ from \eqref{tq} and defining 
\begin{eqnarray}
\Re(u)\equiv Q(u)\exp\Big(-\frac{1}{\eta}\int^{u}a(y)dy\Big)=\prod_{k}(u-\mu_{k})\prod_{n}(u-\xi_{n})^{-s_n},
\end{eqnarray}
we can verify that there holds 
\begin{eqnarray}
\eta x^{\prime}+x^{2}=\tau,~~~~~~~~~x(u)=\eta\frac{\Re^{\prime}(u)}{\Re(u)}= 
-a+\sum_{k}\frac{\eta}{u-\mu_{k}}.
\label{hold}
\end{eqnarray}
This is the so-called Lam\'{e} equation in disguise. Equivalently, 
$\Re(u)$ solves a Fuchs-type equation $\big( \eta^2\partial_u^2 - \tau(u)\big) \Re(u)=0$ with 
$N$ regular singularities on ${\bf{CP}}^1$.

Compared with $x^2$, $\eta x'$ becomes subleading. 
Further getting rid of $\eta x^{\prime}$, we 
arrive at Gaudin's spectral curve
\begin{eqnarray}
 x^{2}=\tau.
\label{LLL}
\end{eqnarray}
In view of \eqref{hold}, it is tempting to introduce $\phi_{KS}$, i.e. Kodaira-Spencer field w.r.t. $Z^M$ defined in \eqref{MMM}. 
That is, 
\begin{eqnarray}
2x\equiv  \partial\phi_{KS}={\cal W}'+2\eta\tr\Big\langle\frac{1}{u-{\cal M}}\Big\rangle_{Z^M}. 
\label{KS}
\end{eqnarray}
Subsequently, \eqref{LLL} becomes precisely the spectral curve 
of $Z^M$. 
Remark that 
\begin{eqnarray}
\oint \partial \phi_{KS}du =-\oint \lambda^\eta_{SW} 
\label{13}
\end{eqnarray} 
up to 
a total derivative term. Additionally, it is well-known that from the 
period integral \eqref{13} one yields 
the tree-level 
free energy ${\cal F}_0$ of $Z^M$:  
\begin{eqnarray}
\label{MMM}
Z^M=\int_{K \times K} {\cal {DM}}\exp\left[{\frac{1}{\eta}\cal{W(M)}}\right],~~~~~~~~~
{\cal W}'=-\sum_{n=1}^{N}
\frac{\ell_n}{(u-\xi_{n})},~~~~~~~~~K\eta=\text{fixed}.
\end{eqnarray}
Of course, the saddle-point of $Z^M$ 
is dictated by \eqref{cb}. We want to display in Sec. 4 
that ${\cal F}_0$ is surely related to 
${\cal T}_{0,N}(A_1)$. In view of \eqref{13}, we refer to this 
as the advertised $SU(N)/SU(2)^{N-3}$ $(N > 3)$ correspondence.

For Gaudin's spectral curve, due to 
$(x,u)\in \bf{C} \times \bf{C}^\ast$ we introduce $v=xu$ such that 
$x du$ here and the former $ud\log w$ of ${\cal T}_{0,4}(A_{N-1})$ look more symmetrical. 
Moreover, the $\pi/2$-rotation noted 
in Fig. \ref{2} is only pictorial otherwise one naively has 
$SU(N)/SU(2)^{N-1}$ correspondence instead%
\footnote{We thank Yuji Tachikawa for his 
comment on this point.}.

\subsubsection{Free-field representation}

As another crucial step, we rewrite 
$Z^M$ in terms of a multi-integral over 
diagonal elements of ${\cal M}$: 
\begin{eqnarray}
Z^M \equiv
 \oint dz_1\cdots \oint d z_K \prod_{i<j}(z_i -z_j)^{2} 
 \prod_{i,n} (z_i-\xi_n)^{-\widetilde{\ell}_n} 
 \prod_{n<m} (\xi_n
 -\xi_m)^{{\widetilde{\ell}_n \widetilde{\ell}_m}/2}.
\end{eqnarray}
A constant term involving only $\xi$'s is multiplied by hand. 
This form then realizes a chiral 
conformal block of $N$ LFT primary fields via 
Feigin-Fuchs free-field representation. 
Notably, the charge balance condition is respected in the presence of background charge $Q$ via inserting 
$K$ screening operators $\oint dz \exp {2}{b^{-1}}\phi(z)$ of 
zero conformal weight. Also, the 
free propagator 
$\big\langle \phi(z_1) \phi(z_2) \big\rangle_{\text{free}}
=-\log (z_1-z_2)^{1/2}$ is used. 

Assume the genus expansion $Z^M=\exp(\eta^{-2}{\cal F}_0+\cdots)$
and 
\begin{eqnarray*}
\lim_{b\to 0}\log \Big\langle 
V_{-\widetilde{\ell}_1/2b}(\xi_1)
\cdots V_{-\widetilde{\ell}_N/2b}(\xi_N) \Big
\rangle_{\text{conformal~block}}=
- b^{-2} \widetilde{F}.
\end{eqnarray*}
$\widetilde{F}$ named $classical$ conformal block appeared in the pioneering work of Zamolodchikov and 
Zamolodchikov \cite{Zamolodchikov:1995aa}. Based on the above discussion, one can anticipate that $\eta^2 \widetilde{F}= {\cal F}_0$. 
Next, to identify ${\cal F}_0$ with the $\Omega$-deformed SW prepotential for ${\cal T}_{0,N}(A_1)$ 
serves as the last step 
towards completing our proposal.

\section{Application and discussion}

Without loss of generality, we examine a concrete 
example: $N=4$. As a result, ${\cal F}_0$ generated by the period integral of $\lambda^\eta_{SW}=-\partial \phi_{KS} du$ is 
indeed the very 
$\epsilon_1$-deformed SW prepotential of 
${\cal T}_{0,4}(A_1)$. Quote the known $\tau(u)$ for $N=4$ 
from 
\cite{Zamolodchikov:1995aa}:
\begin{eqnarray}
\eta^{-2} \tau(u)=\frac{\delta_{1}}{u^{2}}+\frac{\delta_{2}}{(u-q)^{2}}+\frac{\delta_{3}}{(1-u)^{2}}+\frac{\delta_{1}+\delta_{2}+\delta_{3}-\delta_{4}}{u(1-u)}+\frac{q(1-q)\widetilde{c}(q)}{u(u-q)(1-u)}.
\end{eqnarray}
Via projective invariance $q$ represents the cross-ratio of four marked points $(\xi_1,\xi_2,\xi_3,\xi_4)\equiv (0,1,q,\infty)$ on ${\bf{CP}}^1$. 

The residue of $\tau$ around $u=1$ is ($v=xu$)
\begin{eqnarray}
\label{qcq}
&&q\widetilde{c}(q)=
-\eta^{-2}\oint ux^2 du
 = -\frac{1}{2}\eta^{-2}\oint v\lambda_{SW}^{\eta}
=q\frac{\partial}{\partial {q}} {\widetilde{F}}_{\delta,\delta_{n}}(q) ,~~~~~~~~~(n=1, \cdots,4),\\
&& \widetilde{F}_{\delta,\delta_{n}}(q) = 
(\delta-\delta_{1}-\delta_{2})\log q 
+ \frac{(\delta+\delta_{1}-\delta_{2})(\delta+\delta_{3}-\delta_{4})}{2\delta}q 
+ {\cal O}(q^{2})\nonumber
\end{eqnarray}
where Polyakov's conjecture \eqref{polyakov} is applied in the last equality of \eqref{qcq}. 
Notice that only the holomorphic $\widetilde{F}$ in 
${\cal S}_{\text{tot}}$ 
survives ${\partial}/{\partial}q$. Conversely, by taking into account the 
stress-tensor nature of the spectral curve $(\partial \phi_{KS})^2=4\tau$ in Hermitian matrix models, 
$\widetilde{F}$ can be replaced by 
${\cal F}_0$ as a result of Virasoro algebra. This observation supports the above $\eta^2 \widetilde{F}= {\cal F}_0$.

Finally, we need another ingredient: Matone's relation \cite{11,12,13}. As is proposed in \cite{14,Poghossian:2010pn, Fucito:2011pn}, 
the $\epsilon_1$-deformed version is 
\begin{eqnarray}
\langle\tr\Phi^{2} \rangle_{\epsilon_1}=2\bar{q}\partial_{\bar{q}}{W},~~~~~~~~~
\bar{q}=\exp(2\pi i \bar{\tau}_{UV})
\label{ton}
\end{eqnarray}
for, say, ${\cal N}=2$ ${\cal T}_{0,4}(A_1)$ theory where   
\begin{eqnarray*}
&&\frac{1}{\epsilon_1\epsilon_2}
{W(\epsilon_1)}
\equiv\lim_{\epsilon_{2}\to 0}\log Z_{\text{Nek}}\big(a,\vec{m},\bar{q},\epsilon_{1},\epsilon_{2}\big),~~~~~\\
&&{a}:~\text{UV~vev~of~}\Phi,~~~~~~~~~\vec{m}:~
\text{four~bare~flavor~masses}.
\end{eqnarray*}
Now, \eqref{qcq} and \eqref{ton} together manifest $\lambda^\eta_{SW}$ as the 
$\epsilon_1$-deformed SW differential for ${\cal T}_{0,4}(A_1)$ if 
there holds 
\begin{eqnarray}
\frac{1}{b^2}
{\widetilde{F}}_{\delta,\delta_{n}}(q)=
\frac{1}{\hbar^2}{\cal F}_0
=\lim_{\epsilon_{2}\to 0} 
\frac{1}{\epsilon_1\epsilon_2}
{W(\epsilon_1)}
\label{hold1}
\end{eqnarray}
under 
$q=\bar{q}$, $\epsilon_1=\eta$ and $\epsilon_{1} {\epsilon_{2}}=\hbar^2$. In fact, \eqref{hold1} has already been verified in \cite{ta}. 
\begin{table}[tb]
\begin{center}
\begin{tabular}{c}
\hline
Theory of RHS in Fig. \ref{2} ($N=4$) \\ \hline
$q= \dfrac{(\xi_1 -\xi_3)(\xi_2 -\xi_4)}{(\xi_2 -\xi_3)(\xi_1 -\xi_4)}$ \\ \hline
$\epsilon_1 {m} \leftrightarrow \widetilde{\ell}$  \\ \hline
$(\epsilon_1, \epsilon_2)=(\eta, 0)$ \\ \hline
$a=\oint_\alpha  \lambda^\eta_{SW}$ \\ \hline 
$\epsilon_1^2 \dfrac{\partial \widetilde{F}}{\partial a}=\oint_\beta  
\lambda^\eta_{SW}$  \\ \hline
\end{tabular}
\caption{Quantities of RHS in Fig. \ref{2} in terms of spin-chain variables}
\label{tab3}
\end{center}
\end{table}
To conclude, by examining 
$\lambda^\eta_{SW}$ we have found that Baxter's T-Q equation 
encodes simultaneously 
two kinds of ${\cal N}=2$ theories, ${\cal T}_{0,N}(A_{1})$ and ${\cal T}_{0,4}(A_{N-1})$. 
We call this remarkable property $SU(N)/SU(2)^{N-3}$ correspondence.

\subsection{Discussion}

$\bullet $ Based on 
\eqref{new} and \eqref{swd}, we have at the level of $\lambda^\eta_{SW}$
\begin{eqnarray} 
\log w=2\eta \frac{\Psi'}{\Psi},~~~~~~~~~
w=\frac{A + \sqrt{A^2-4}}{2},~~~~~~~~~A =\frac{P_{N_c}}{\sqrt{Q_{N_f}}}.
\end{eqnarray}
Namely, all quantum $SU(N_c)$ Coulomb moduli encoded inside $P_{N_c}(u)\equiv \langle \det(u-\Phi) \rangle$ are determined by using spin-chain variables $(\eta, \xi , \ell)$. 
This fact is consistent with 
\eqref{phypre}. \\
$\bullet $ Besides, from Table \ref{tab3} we find that the transformation between 
${\cal F}_{SW}$ in \eqref{phypre} and $\widetilde{F}$ is quite complicated. Although sharing 
the same SW differential (up to a total derivative term), two theories have diverse 
IR dynamics because both of their gauge group and matter content differ. 
To pursue a concrete interpolation between them is under investigation.

\subsection{XYZ Gaudin model}
There are still two other Gaudin models, 
say, trigonometric and elliptic ones. 
Let us briefly discuss the elliptic type because it sheds light on $\N=2$ ${\cal T}_{1,n}(A_{1})$ theory. 
Now, Bethe roots satisfy the following classical Bethe Ansatz equation:
\begin{eqnarray} 
\sum_{n=1}^{N}\frac{s_{n}\theta_{11}^{\prime}(\mu_{k}-\xi_{n})}{\theta_{11}(\mu_{k}-\xi_{n})}=
-\pi i\nu+\sum_{l(\ne k)}\frac{\theta_{11}^{\prime}(\mu_{k}-\mu_{l})}{\theta_{11}(\mu_{k}-\mu_{l})},~~~~~~~~~\nu\in\text{integer}.\label{ecb}
\end{eqnarray}

Regarding it as a saddle-point condition, we are led to 
the spectral curve analogous to \eqref{LLL} 
\begin{align*}
\begin{split}
x^2&=\left[\displaystyle \sum_{n=1}^{N}\frac{s_{n}\theta_{11}^{\prime}(u-\xi_{n})}{\theta_{11}(u-\xi_{n})}-
\sum^K_{k=1}\frac{\theta_{11}^{\prime}(u-\mu_{k})}{\theta_{11}(u-\mu_{k})}\right]^{2}\\ 
&=
\sum_{n=1}^{N} s_{n}  (s_{n}+1)\wp (u-\xi_n) +\sum_{n=1}^{N}H_{n}\zeta(u-\xi_{n})+H_{0}
\end{split}
\end{align*}
where
\begin{align}
\begin{split}
H_n&=\sum^N_{i\ne n} 
\sum_{a=1}^3 w_a(\xi_n - \xi_i)
J_n^a
J_i^a,\\
H_0 &= \sum_{n=1}^N \sum_{a=1}^3 \Bigl\{
            -  \wp\left(\frac{\omega_{5-a}}{2}\right)J_n^a J_n^a +
              \sum_{i \neq n} w_a(\xi_i -\xi_n)
              \left[
              \zeta\Bigl(\xi_n - \xi_i +\frac{\omega_{5-a }}{2}\Bigr)
            - \zeta\Bigl(\frac{\omega_{5-a }}{2}\Bigr)
              \right] J_n^a J_i^a
            \Bigr\}.
\end{split}\label{eq:XYZH}
\end{align}
Here, $\wp(u)$ and $\zeta(u)$ respectively denote Weierstrass $\wp$- and $\zeta$-function.
Periods of $\wp(u)$ are (see Appendix \ref{Appendix_elliptic} for $w_a$)
\begin{align}
\begin{split}
\omega_1=\omega_4=1,~~~~~~~~~
\omega_2=\tau,~~~~~~~~~
\omega_3=\tau+1.
\end{split}\label{eq:periods}
\end{align}
Notice that $H_n$'s ($\sum H_n =0$) are known as elliptic Gaudin Hamiltonians \cite{Sklyanin_Takebe1996, Sklyanin_Takebe1999}. 
All these are elliptic counterparts of those in the rational XXX model.  According to the logic of Fig. \ref{3}, it will be 
interesting to verify whether the XYZ one-form 
$ x du $ reproduces the $\epsilon_1$-deformed $\N=2^\ast$ SW 
prepotential when $n=1$%
\footnote{See also \cite{Alday:2010vg,Maruyoshi:2010iu} for related discussions.}

\section*{Acknowledgments}
TST thanks Kazuhiro Sakai, Hirotaka Irie and Kohei Motegi for encouragement and helpful comments. 
RY is supported in part by Grant-in-Aid 
for Scientific Research No.23540316 from Japan Ministry of Education. 
NY and RY are also supported 
in part by JSPS Bilateral Joint Projects (JSPS-RFBR collaboration).

\appendix

\section{Definition of $w_n$}
\label{Appendix_elliptic}
In Appendix \ref{Appendix_elliptic}, $w_n$ that appears in (\ref{eq:XYZH}) 
will be defined according to \cite{Sklyanin_Takebe1996,Sklyanin_Takebe1999}.
We choose periods of $\wp(u)$ as in (\ref{eq:periods}).
Weierstrass $\sigma$-function is as follows: 
\begin{align}
\begin{split}
\sigma(u)&=\sigma(u;\omega_1,\omega_2)\\
&=
u\prod_{
    \begin{subarray}{c}
        (n,m) \neq (0,0)\\
        n,m \in \bf{Z}
    \end{subarray} 
}
\left(1-\frac{u}{n\omega_1+m\omega_2}\right)
\exp\left[
\frac{u}{n\omega_1+m\omega_2}+\frac{1}{2}\left(\frac{u}{n\omega_1+m\omega_2}\right)^2
\right].
\end{split}
\end{align}
Note that $\sigma(u)$ satisfies 
\begin{align}
\begin{split}
\zeta(u)=
\frac{\sigma'(u)}{\sigma(u)},~~~~~~~~~~~~
\wp(u)=
-\zeta'(u).
\end{split}
\end{align}
We introduce $(e_a, \eta_a, \varsigma_a)$ which are 
related to $\omega_a$ by  
\begin{align}
\begin{split}
e_a=\wp(\omega_a/2),~~~~~~~
\eta_a=\zeta(\omega_a/2),~~~~~~~
\varsigma_a=\sigma(\omega_a/2),~~~~~~a=1,2,3.
\end{split}
\end{align}
Using them we further have 
\begin{align}
\begin{split}
\sigma_{00}(u)
&=\frac{\exp\left[
        -\left(\eta_1+\eta_2\right)u
    \right]}{\varsigma_3}
    \sigma\left(u+\frac{\omega_3 }{2}\right),\\
\sigma_{10}(u)
&=\frac{\exp\left(
        -\eta_1 u
    \right)}{\varsigma_1}
        \sigma\left(u+\frac{\omega_1}{2}\right),\\
\sigma_{01}(u)
&=\frac{\exp\left(
        - \eta_2 u
    \right)}{\varsigma_2}
        \sigma\left(u+\frac{\omega_2}{2}\right).
\end{split}
\end{align}
Note that Jacobi's $\vartheta$-functions are 
\begin{align}
\begin{split}
\vartheta_{00} (u) 
&= \vartheta(u;\tau)=\vartheta(u)=\sum_{n=-\infty}^{\infty}\exp \left(\pi {i} n^2 \tau + 2 \pi {i} n u\right),\\
\vartheta_{01} (u) 
&= \vartheta \left(u+\frac{1}{2}\right),\\
\vartheta_{10} (u) 
&= \exp\left(\frac{1}{4}\pi{i}\tau+\pi{i} u\right)
    \vartheta \left(u+\frac{1}{2}\tau\right),\\
\vartheta_{11} (u) 
&= \exp\left(\frac{1}{4}\pi{i}\tau+\pi{i}(u+\frac{1}{2})\right)
    \vartheta \left(u+\frac{1}{2}+\frac{1}{2}\tau\right)
\end{split}
\end{align}
from which Weierstrass $\sigma$-functions are defined as below:
\begin{align}
\begin{split}
\omega_1
\exp\left(
\frac{\eta_1}{\omega_1}u^2
\right)
\frac{
        \displaystyle\vartheta_{11}\left(\frac{u}{\omega_1}\right)
}{
        \displaystyle
        \vartheta'_{11}\left(0\right)
}=\sigma(u),~~~~~~
\exp\left(
\frac{\eta_1}{\omega_1}u^2
\right)
\frac{
        \displaystyle\vartheta_{ab}\left(\frac{u}{\omega_1}\right)
}{
        \displaystyle
        \vartheta'_{ab}\left(0\right)
}=\sigma_{ab}(u)~~~~~~(ab=0).
\end{split}
\end{align}
Finally, $w_a(u)$ can be obtained as follows: 
\begin{align}
\begin{split}
    w_1(u) &= \frac{\mathrm{cn}(u\sqrt{e_1-e_3};\sqrt{\dfrac{e_2-e_3}{e_1-e_3}})}{\mathrm{sn}(u\sqrt{e_1-e_3};\sqrt{\dfrac{e_2-e_3}{e_1-e_3}})}
        =\frac{\sigma_{10}(u)}{\sigma(u)}    = \frac{\vartheta'_{11}(0)}{\vartheta_{10}(0)}
      \frac{\vartheta_{10}(u)}{\vartheta_{11}(u)},~~~~~
\\
    w_2(u) &=\frac{\mathrm{dn}(u\sqrt{e_1-e_3};\sqrt{\dfrac{e_2-e_3}{e_1-e_3}})}{\mathrm{sn}(u\sqrt{e_1-e_3};\sqrt{\dfrac{e_2-e_3}{e_1-e_3}})}
        =\frac{\sigma_{00}(u)}{\sigma(u)}    = \frac{\vartheta'_{11}(0)}{\vartheta_{00}(0)}
      \frac{\vartheta_{00}(u)}{\vartheta_{11}(u)},
\\
    w_3(u) &=\frac{1}{\mathrm{sn}(u\sqrt{e_1-e_3};\sqrt{\dfrac{e_2-e_3}{e_1-e_3}})}
        =\frac{\sigma_{01}(u)}{\sigma(u)}= \frac{\vartheta'_{11}(0)}{\vartheta_{01}(0)}
      \frac{\vartheta_{01}(u)}{\vartheta_{11}(u)}.
\end{split}
\end{align}

\end{document}